# Visual Analytics for Explainable Deep Learning

**Jaegul Choo**
Korea University

**Shixia Liu**
Tsinghua University

Recently, deep learning has been advancing the state of the art in artificial intelligence to a new level, and humans rely on artificial intelligence techniques more than ever. However, even with such unprecedented advancements, the lack of explanation regarding the decisions made by deep learning models and absence of control over their internal processes act as major drawbacks in critical decision-making processes, such as precision medicine and law enforcement. In response, efforts are being made to make deep learning interpretable and controllable by humans. In this paper, we review visual analytics, information visualization, and machine learning perspectives relevant to this aim, and discuss potential challenges and future research directions.

## WHY EXPLAINABLE DEEP LEARNING?

Deep learning has had a considerable impact on various long-running artificial intelligence problems, including computer vision, speech recognition and synthesis, and natural language understanding and generation [1]. As humans rely on artificial intelligence techniques, the interpretability of their decisions and control over their internal processes are becoming a serious concern for various high-impact tasks, such as precision medicine, law enforcement, and financial investment. Gender and racial biases learnt by artificial intelligence programs recently emerged as a serious issue.[1] In April 2016, the European Union legislated the human right to request an explanation regarding machine-generated decisions.[2]

Interpretation is the process of generating human-understandable explanations on why a particular decision is made by a deep learning model. However, the end-to-end learning paradigm hides the entire decision process behind the complicated inner-workings of deep learning models, making it difficult to obtain interpretations and explanations.

Recently, considerable effort has been invested to tackle this issue. For instance, Defense Advanced Research Projects Agency (DARPA) of the United States is launching a large initiative called Explainable Artificial Intelligence (XAI).[3] Machine learning and artificial intelligence communities have conducted many relevant workshops and meetings.[4]

[1] https://www.wired.com/story/machines-taught-by-photos-learn-a-sexist-view-of-women/

[2] https://www.wired.com/2016/07/artificial-intelligence-setting-internet-huge-clash-europe/

Interactive visualization plays a critical role in enhancing the interpretability of deep learning models, and it is emerging as a promising research field. Recently, in premier venues in this field such as IEEE VIS, a growing number of papers concerning interactive visualization and visual analytics for deep learning have been published, and the best paper in IEEE VIS'17 has been awarded to a visual analytics system that supports advanced interactive visual capabilities in TensorBoard, a visualization tool for Google's TensorFlow [2].

Given the deluge of deep learning techniques and their applications, this article provides a comprehensive overview of recent research concerning interpretability and explainability of deep learning using basic toolkits, advanced algorithmic techniques, and intuitive and interactive visual interfaces. Based on a systematic analysis of the aforementioned research, we describe current research challenges and promising future directions.

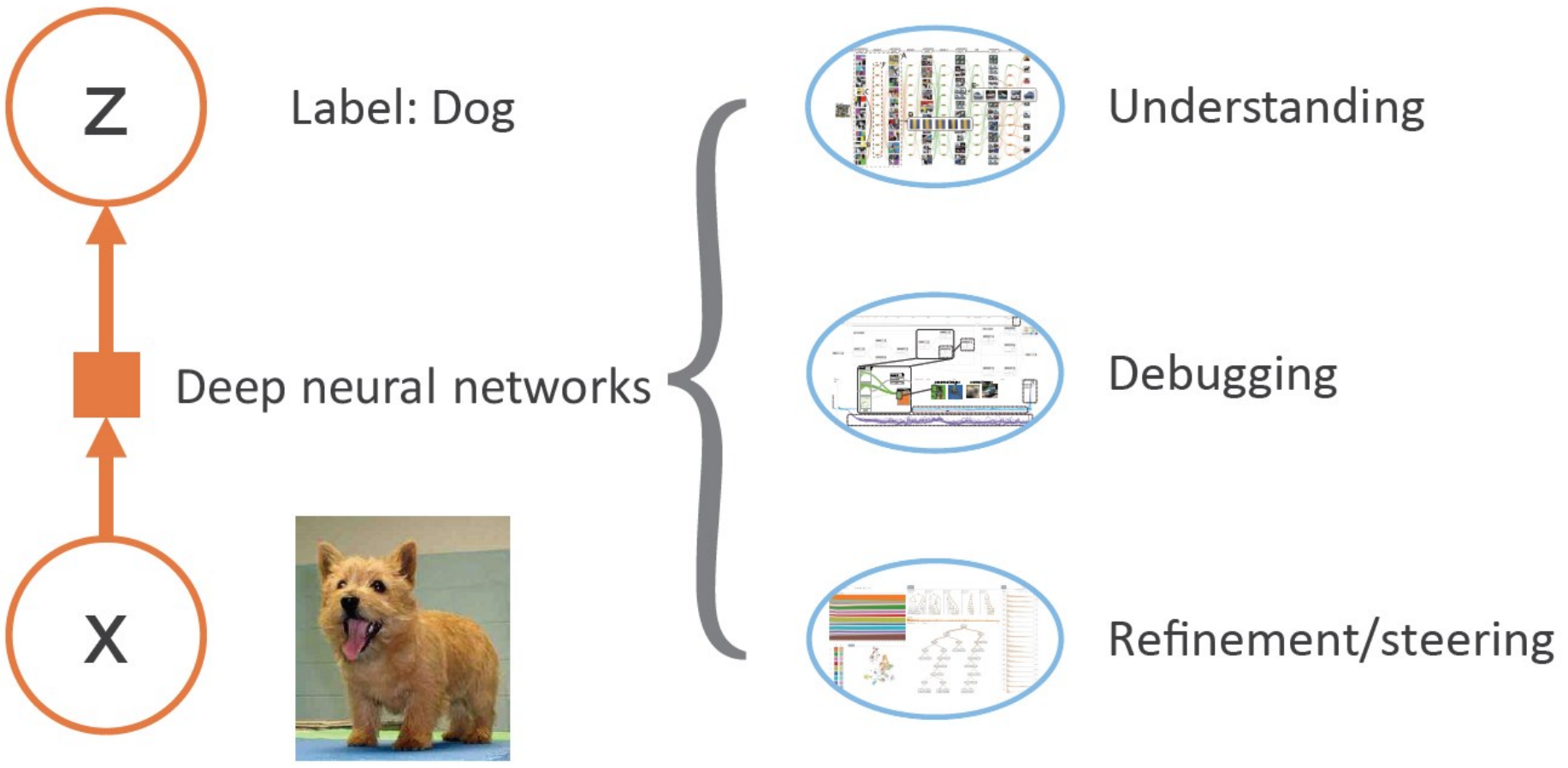

Figure 1. Overview of explainable deep learning

# OVERVIEW OF EXPLAINABLE DEEP LEARNING

As shown in Figure 1, explainable deep learning encompasses three major research directions: model understanding, debugging, and refinement/steering. Model **understanding** aims to explain the rationale behind model predictions and the inner workings of deep learning models, and it attempts to make these complex models at least partly understandable. Model **debugging** is the process of identifying and addressing defects or issues within a deep learning model that fails to converge or does not achieve an acceptable performance [9]. Model **refinement/steering** is a method to interactively incorporate expert knowledge and expertise into the improvement and refinement process of a deep learning model, through a set of rich user interactions, in addition to semi-supervised learning or active learning. In the following, we will discuss recent approaches to explainable deep learning in these directions.

# EDUCATIONAL USE AND INTUITIVE UNDERSTANDING WITH INTERACTIVE VISUALIZATION

Interactive visualization has played an important role in providing an in-depth understanding of how deep learning models work. Tensorflow Playground[5] is an effective system for education and intuitive understanding, where users can play with simple neural networks by changing various configurations in terms of the numbers of layers and nodes, and the types of nonlinear units. It adopts two-dimensional toy data sets for classification and regression tasks. The manner in which each node in the network is activated across different input data values is fully visualized as a heatmap in a two-dimensional space.

[3] https://www.darpa.mil/program/explainable-artificial-intelligence

[4] A partial list of workshops can be found at http://icmlviz.github.io/reference/

Another web-based deep learning library called ConvNetJS[6] features easy access to deep learning techniques by simply using a web browser, together with a rich set of visualization modules, which can be effectively used for education and understanding. DeepVis toolbox[7] dynamically visualizes the activation map of various filters in a user-selected layer of convolutional neural networks (CNNs), given a webcam video input in real time.

These tools and systems provide effective interactive visualization for interpreting deep learning models, but most of them are limited to simple models and basic applications, and thus their applicability remains far from real-world problems.

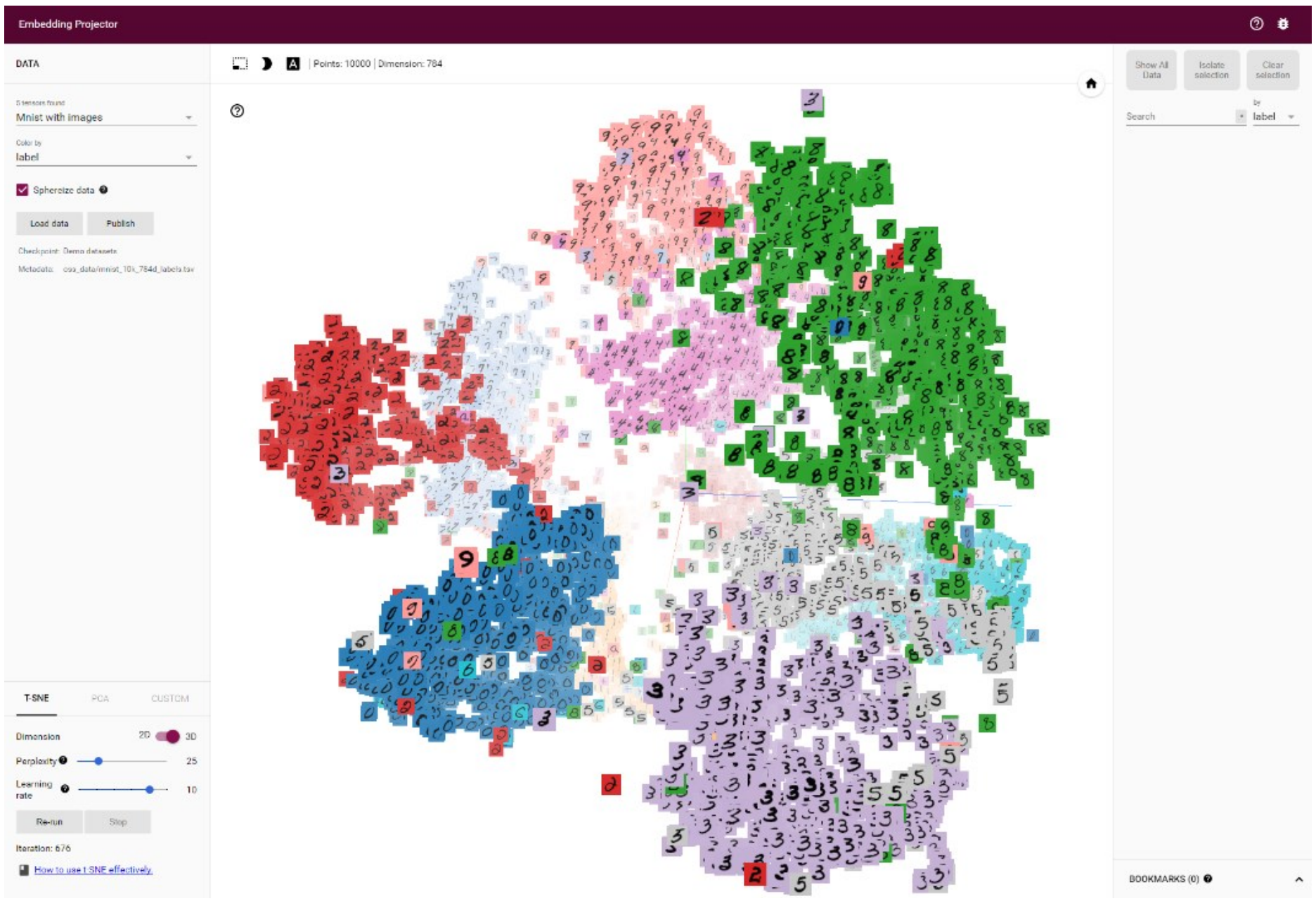

Figure 2. Embedding projector based on 2D/3D embedding techniques, such as principal component analysis and *t*-distributed stochastic embedding. In this example, MNIST handwritten images are visualized as rectangles colored by their associated digit labels so that those images with high similarity in their original feature space are placed close to each other in the 2D/3D space. In this manner, one can easily identify which digit clusters are similar (and thus confusing from a classifier's perspective) and which digit images are outliers (and thus confusing as another digit).

## MODEL DEBUGGING THROUGH VISUALIZATION TOOLKITS

Most deep learning libraries are often accompanied by basic visualization toolkits that allow users to debug current models and improve performance. For example, TensorFlow's TensorBoard visualizes the structure of a given computational graph that a user creates and provides basic line graphs and histograms of user-selected statistics, such as the loss value and activation and/or gradient value of a particular node. Recently, as shown in Figure 2, TensorFlow has been equipped with a new visualization module called Embedding Projector.[8] This module provides a 2D/3D embedding view using principal component analysis and *t*-distributed stochastic neighbor embedding, which reveals the relationships between data points with respect to their multi-dimensional representations in a given layer.

[5] http://playground.tensorflow.org

[6] https://cs.stanford.edu/people/karpathy/convnetjs/

[7] http://yosinski.com/deepvis

Visdom[9] is a web-based interactive visualization toolkit that is easy to use with deep learning libraries such as PyTorch. Deeplearning4j UI[10] is another visual user interface, which allows users to monitor the training process with several basic visualization components.

Although these visualization toolkits offer an intuitive presentation of the low-level information directly provided by deep learning models, it remains difficult for humans to understand the behaviors of these models at a semantically meaningful level.

## COMPUTATIONAL METHODS FOR INTERPRETATION AND EXPLANATION

In the academic field of machine learning, the interpretation of a deep learning model typically refers to the task of identifying the feature importance score, e.g., which part of the input feature of a given data item is responsible for the prediction result at the output layer and/or the high activation of an internal layer/node.

The machine learning and artificial intelligence communities have been developing new techniques to solve this problem. Perturbation experiments [3] and saliency map-based methods [4] have shown their capabilities in revealing which part of the input image is most responsible for the final prediction of a model. A recently proposed technique called LIME [5] approximately constructs a locally linear model from a complex model, in order to allow the interpretation of linear combination weights as the feature importance score.

Furthermore, the training data items most responsible for a particular prediction output have been identified in an efficient manner by adopting a classical technique called influence functions [6]. The corresponding paper was selected for the best paper award in ICML 2017.

In fact, the integration of these advanced computational methods with an interactive visualization can bring significant potentials towards explainable deep learning and remains a major challenge in real-world applications.

## VISUAL ANALYTICS FOR IN-DEPTH UNDERSTANDING AND MODEL REFINEMENT

The visual analytics and information visualization communities have recently developed intuitive and interactive user interfaces with advanced visualization modules. Such visual analytics systems provide users with an in-depth understanding and clues concerning how to troubleshoot and further improve a given model [7,8].

---

[8] http://projector.tensorflow.org/

[9] https://github.com/facebookresearch/visdom

[10] https://deeplearning4j.org/visualization

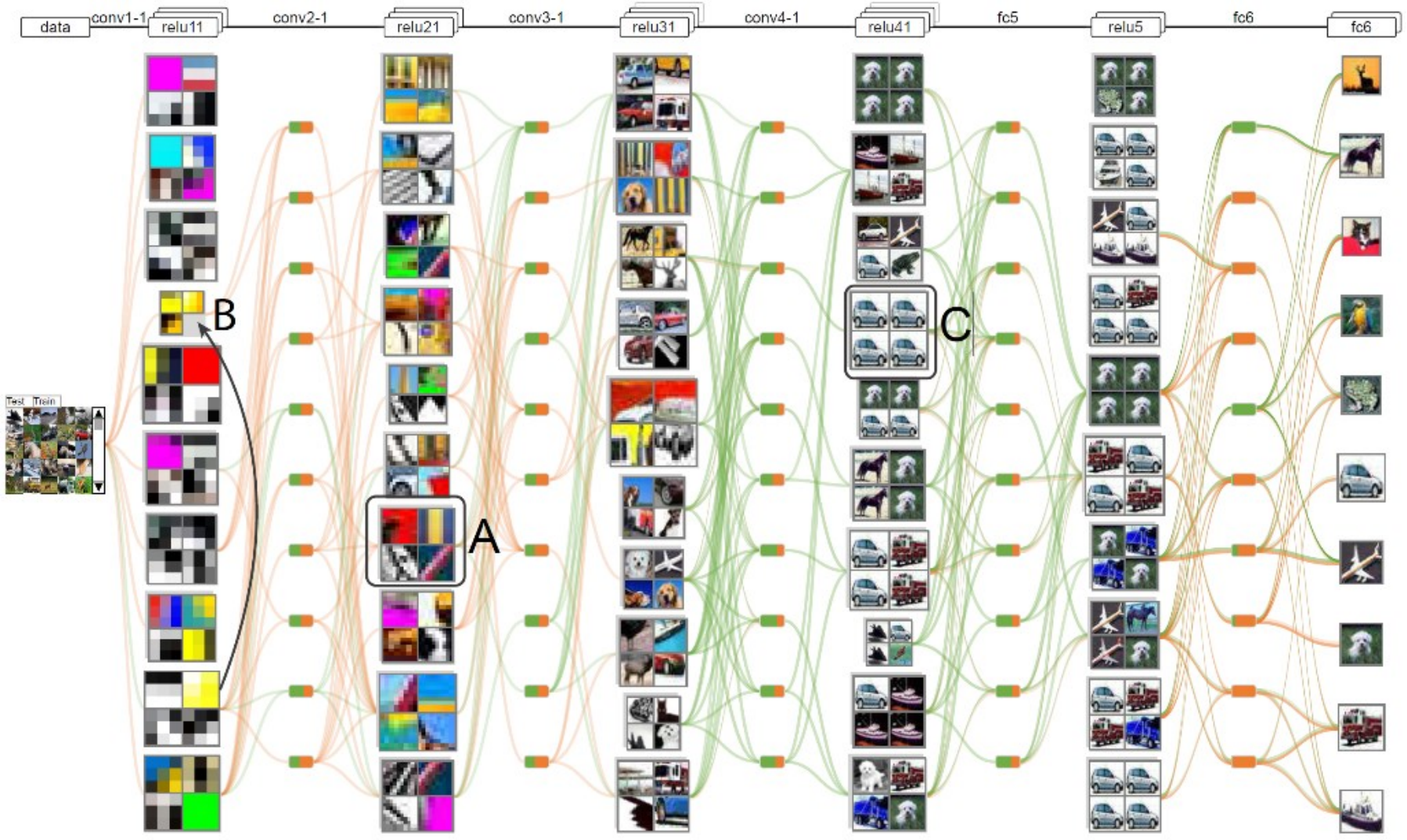

Figure 3. Overview of CNNVis [8]. The system employs a DAG to visually illustrate the inner workings of a CNN training process. It also performs bi-clustering to visualize the strength of interactions between filters of adjacent layers, along with their most activated images.

CNNVis [8] is a representative example of a visual analytics system for model understanding and the diagnosis of CNNs. As shown in Figure 3, by leveraging the DAG layout and examining the learned representations, CNNVis investigates how the depth and width of neural networks influence the features detected by the neurons and the model classification performance. It also helps machine learning experts to diagnose a failed training process. ActiVis [10] provides a visual exploratory analysis of a given deep learning model via multiple coordinated views, such as a matrix view and an embedding view.

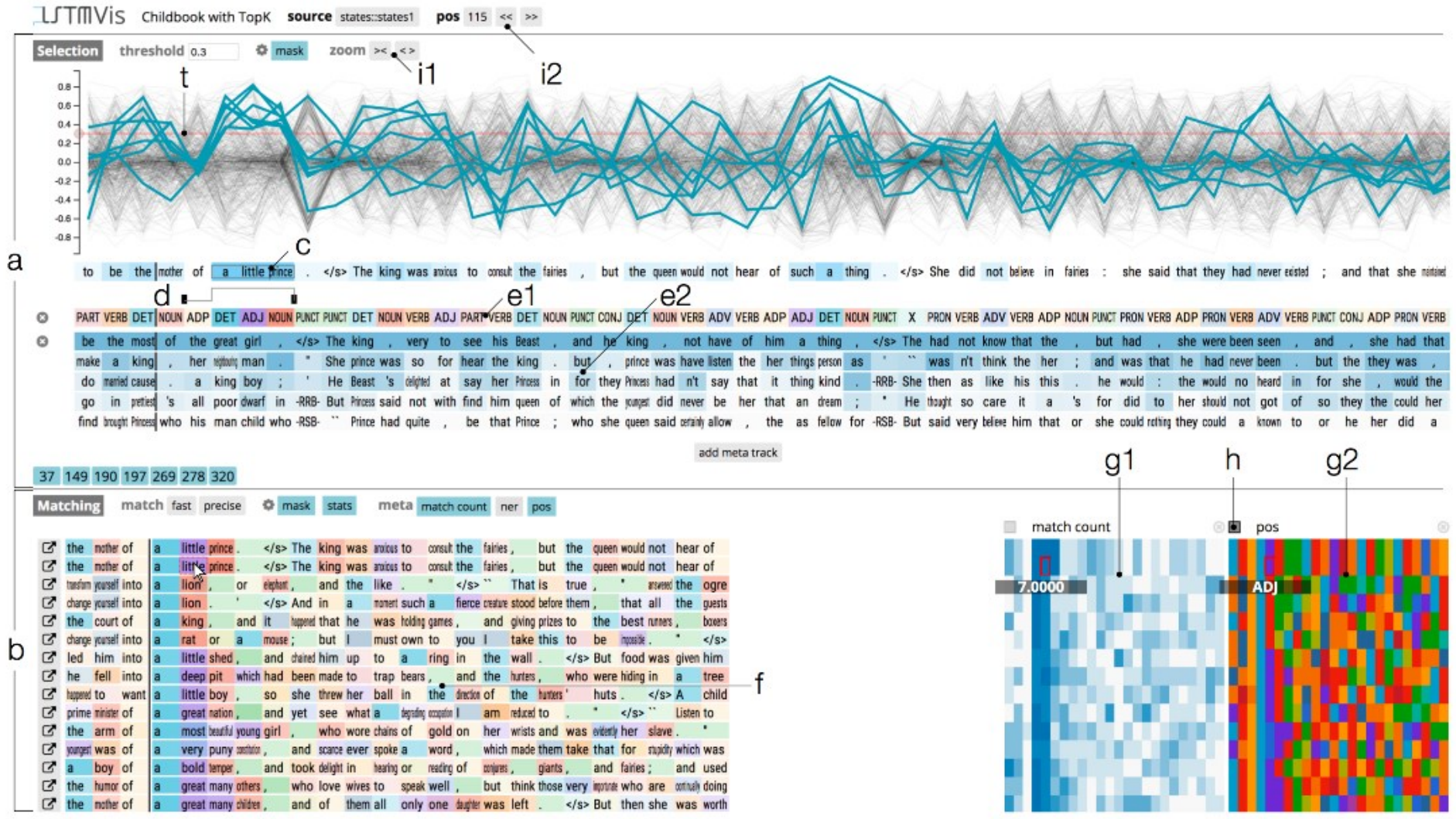

Figure 4. Overview of LSTMVis [10]. Given a sequential input, e.g., words in a document, the line graphs in the top panel visualize the activation patterns of individual hidden nodes over the sequence. Once a user selects a sub-sequence containing multiple words, e.g., the noun phrase "a little prince," and specifies a threshold, the system identifies those hidden nodes exhibiting activation values greater than the threshold and finds other sub-sequence examples for which the same hidden nodes have high activation values. These sub-sequences allow users to interactively explore the learnt behavior of hidden nodes.

Some initial effort has been made to understand recurrent neural networks (RNNs) and their widely employed architecture called long short-term memory (LSTM). For example, in order to provide an idea of the semantic meanings of cells in language modeling applications, LSTMVis [11] presents the activation patterns of individual cells over time steps or sequences as line graphs (Figure 4). RNNVis [12] performs clustering on hidden-state nodes with similar activation patterns and visualizes them as grid-style heatmaps, along with their most strongly associated keywords.

There also exist visual analytics systems that visualize the training processes of deep learning models in real time and facilitate the debugging/improvement of model accuracy and computational time. ReVACNN [13] is one such visual analytics system for CNNs, which provides real-time model steering capabilities during training, such as dynamically removing/adding nodes and interactively selecting data items for a subsequent mini-batch in the training process. In addition, DeepEyes [14] is capable of real-time monitoring and interactive model steering of deep learning models, for example by highlighting stable nodes and layers. DeepEyes has also made some initial effort towards model refinement, by allowing users to remove filters with very low activation values. DGMTracker employs the blue noise sampling algorithm and credit assignment algorithm to detect which portions of the input images cause a training failure for a particular image set in deep generative models, such as generative adversarial networks and variational autoencoders [15].

Although recent visual analytics systems provide sophisticated visualization and interaction capabilities, research issues on how to effectively loop human into the analysis process and how to increase applicability of explainable deep learning techniques have not been fully investigated. For example, widening user interaction capabilities with deep learning models based on various user needs, tightly integrating the current data-driven learning process with knowledge-driven analysis processes, evaluating and improving the robustness of deep learning against out-of-sample data, as well as explaining other types of popular deep learning models, still pose considerable challenges for explainable deep learning.

# RESEARCH GAPS AND OPPORTUNITIES

The development of a highly accurate and efficient deep learning model is an iterative and progressive process of training, evaluation, and refinement, which typically relies on a time-consuming trial-and-error procedure where the parameters and the model structures are adjusted based on user expertise. Typically, machine learning researchers tend to build a new learning process to trace a model's prediction [5]. For example, model predictions can be explained by highlighted image regions. Visualization researchers are making initial attempts to visually illustrate intuitive model behaviors and debug the training processes of widely-used deep learning models such as CNNs and RNNs [7,11]. However, little effort has been made in tightly integrating state-of-the-art deep learning models/methods with interactive visualizations to maximize the value of both. Based on this gap and our understanding of current practices, we identify the following research opportunities.

## Opportunity 1. Injecting external human knowledge

Currently, most deep learning models constitute data-driven methods, whereas knowledge-driven perspectives have received comparatively little attention. In this sense, an open research opportunity is to combine human expert knowledge and deep learning techniques through interactive visualization. To be specific, potential research topics include domain knowledge representation and interpretation, expert knowledge prorogation, and knowledge-based visual explanation. In addition, visual analytics could be utilized to intuitively verify that a model correctly follows human-injected knowledge and rules, which is a crucial step in ensuring the proper behavior of critical deep learning applications. For example, when training an autonomous driving model with video camera inputs, one can impose the rule of never hitting a person recognized in the scene.

## Opportunity 2. Progressive visual analytics of deep learning

Most of the existing explainable deep learning approaches mainly focus on understanding and analyzing model predictions or the training process offline after the model training is complete. As the training of many deep learning models is time-consuming (requiring from hours to days of computation), progressive visual analytics techniques are needed to incorporate experts into the analysis loop. To this end, deep learning models are expected to produce semantically meaningful partial results during the training process. Experts can then leverage interactive visualizations to explore these partial results, examine newly incoming results, and perform new rounds of exploratory analysis without having to wait for the entire training process to be completed.

## Opportunity 3. User-driven generative models

Traditional classification or regression problems associate each data item with a unique answer in a simple form, and thus the prediction output of deep learning models has little room for human intervention. However, generative models, which perform the generation and translation of, say, images, sentences, and speech signals, can have multiple possible answers, which leaves ample room for the interactive steering of deep learning outputs.

For instance, users can steer the colorization process of a given gray-scale image [16]. One can also transform a given facial image by interactively changing its attributes such as gender and facial expression [17]. Promising directions would be to develop (1) new deep learning models that can flexibly take various user inputs and reflect them in the generative output and (2) novel visualization-based interfaces that allow users to effectively interact with deep learning.

## Opportunity 4. Improving the robustness of deep learning for secure artificial intelligence

Deep learning models are generally vulnerable to adversarial perturbations, where adversarial examples are maliciously generated to mislead the model to output wrong predictions. An adversarial example is modified very slightly, and thus in many cases these modifications can be so subtle that a human observer cannot even notice the modification at all, yet the model still makes a mistake. These adversarial examples are often used to attack a deep learning model.

In this respect, maintaining the robustness of a deep learning model is critical in real-world applications. Accordingly, one research opportunity concerning explainable deep learning is to incorporate human knowledge to improve the robustness of deep learning models.

## Opportunity 5. Reducing the size of the required training set

Typically, a deep learning model contains hundreds of parameters. To fully train a learning model with such a large number of parameters, thousands of training examples are required. In real-world applications, a method is impractical if each specific task requires its own separate large-scale collection of training examples. To close the gap between academic research outputs and real-world requirements, it is necessary to reduce the sizes of required training sets by leveraging prior knowledge obtained from previously trained models in similar categories, as well as human expertise.

One-shot learning or zero-shot learning [18], which are major unsolved problems in the current practice of training deep learning models, provide a possibility to incorporate prior knowledge on objects into a “prior” probability density function. That is, those models trained using given data and their labels can usually solve only pre-defined problems for which they were originally trained. For example, a deep learning model that detects a cat cannot in principle detect a tiger without sufficient training data with tiger labels. The injection of a small amount of user inputs can potentially solve these problems through a visual analytics framework.

As a result, an interesting direction for future research would be to study how to combine visual analytics with learning capabilities such as one-shot learning or zero-shot learning to incorporate external human knowledge and reduce the number of training samples needed.

### Opportunity 6. Visual analytics for advanced deep learning architectures

So far, researchers have mostly developed visual analytic approaches for basic deep learning architectures, such as CNNs and RNNs. Attention-based models have exhibited their advantages in various applications. Many other advanced architectures have recently been proposed as effective alternatives, and some of these are being effectively employed for state-of-the-art performance in existing tasks (ResNet and DenseNet for image recognition), as well as for new challenging tasks (memory networks and co-attention models for natural-language question answering).

In general, these models are usually not only composed of numerous layers (e.g., several hundreds or thousands in ResNet and DenseNet), but also involve complicated network designs for individual layers, as well as a heavily connected structure between them. Such complexity of advanced deep learning architectures poses unprecedented interpretation and interaction challenges in the visual analytic and information visualization communities.

Accordingly, future research should address the aforementioned issues by developing effective and efficient visualization techniques and the intuitive summarization of such large-scale networks in terms of the number of nodes, layers, and their connectivity. Furthermore, novel interaction techniques should be developed to enhance their interpretability and model-steering capabilities.

## CONCLUSION

Deep learning and artificial intelligence have had a significant impact on our lives, and as we rely heavily on them, explainability and a deep understanding regarding their decisions and internal processes are becoming crucial. In this paper, we reviewed recent efforts from visual analytics, information visualization, and machine learning perspectives in both academia and industry, including basic toolkits, advanced computational techniques, and intuitive and interactive visual analytics systems.

Finally, we discussed the gaps in research, and proposed potential research opportunities such as human-in-the-loop visual analytics integrating human knowledge and data-driven learning approach, progressive visual analytics for deep learning, user-driven generative models, and visual analytics for secure deep learning.

We hope that these proposed directions will inspire new research that can improve the current state of the art in deep learning toward accurate, interpretable, efficient, and secure artificial intelligence.

## ACKNOWLEDGMENTS

We greatly appreciate the feedback from anonymous reviewers. This work was partially supported by the Basic Science Research Program through the National Research Foundation of Korea (NRF) grant funded by the Korea government (MSIP) (No. NRF-2016R1C1B2015924) and National NSF of China (No. 61672308). Any opinions, findings, and conclusions or recommendations expressed here are those of the authors and do not necessarily reflect the views of funding agencies.

# AUTHORS BIOS

**Jaegul Choo** is an assistant professor in the Dept. of Computer Science and Engineering at Korea University. He has been a research scientist at Georgia Tech from 2011 to 2015, where he also received an M.S in 2009 and Ph.D in 2013. His research focuses on visual analytics for machine learning and deep learning. He earned the Best Student Paper at ICDM’16, the Outstanding Research Scientist Award at Georgia Tech in 2015, and the Best Poster Award at IEEE VAST (as part of IEEE VIS) in 2014, and he was nominated as one of the four finalists for the IEEE Visualization Pioneers Group dissertation award in 2013. He can be contacted at jchoo@korea.ac.kr.

**Shixia Liu** Shixia Liu is an associate professor in the School of Software, Tsinghua University. Her research interests include visual text analytics, visual social analytics, visual model analytics, and text mining. Shixia is the associate editor of IEEE Transactions on Visualization and Computer Graphics and on the editorial board of Information Visualization. She was the papers co-chair of IEEE VAST 2016 and 2017 and was the program co-chair of PacifcVis 2014 and VINCI 2012. She can be contacted at shixia@tsinghua.edu.cn.

Contact department editor Theresa-Marie Rhyne at theresamarierhyne@gmail.com.